\documentclass[epj]{svjour}
\usepackage{graphics}
\usepackage{amsmath}
\usepackage{amssymb}

\newcommand{\bra}[1]{\langle #1 | \,}
\newcommand{\ket}[1]{\, | #1 \rangle}

\newcommand{\be}{\begin{equation}}
\newcommand{\ee}{\end{equation}}
\newcommand{\bea}{\begin{eqnarray}}
\newcommand{\eea}{\end{eqnarray}}

\begin{document}
\title{Postponement of dark-count effects in practical quantum key-distribution by 
two-way post-processing}

\author{Aeysha Khalique, Georgios M. Nikolopoulos, \and Gernot Alber} 
%
\offprints{}          
\institute{Institut f\"ur Angewandte Physik, Technische 
Universit\"at Darmstadt, 64289 Darmstadt, Germany}
\date{Received: date / Revised version: \today}
%
\abstract{
The influence of imperfections on 
achievable secret-key generation rates of quantum key distribution protocols 
is investigated. As examples of relevant imperfections, we consider 
tagging of Alice's qubits and dark counts at Bob's detectors, while 
we focus on a powerful eavesdropping strategy which takes full advantage of 
tagged signals.  
It is demonstrated that error correction and privacy amplification based on 
a combination of a two-way classical communication protocol and 
asymmetric Calderbank-Shor-Steane codes may significantly postpone the 
disastrous influence of dark counts. As a result, the 
distances are increased considerably 
over which a secret key can be distributed in optical fibres  reliably.
Results are presented for the four-state, the six-state, and
the decoy-state protocols.
\PACS{
      {03.67.Dd}{Quantum Cryptography}   \and
      {03.67.Hk}{Quantum Communication}
     } 
} 

\authorrunning{A. Khalique {\em et al.}}
\titlerunning{Postponement of dark-count effects in practical QKD by two-way 
post-processing}

\maketitle

\section{Introduction} 
\label{Sec1}
The unconditional security of the four-state \cite{BB84} 
and the six-state \cite{Bruss6} quantum-key-distribution (QKD) 
protocols has been addressed by many authors 
(see e.g., \cite{SP,KP,L01,May01}). 
Although such 
security proofs allow for 
the most general eavesdropping attacks consistent with 
quantum theory (so-called 
coherent or joint attacks), they impose certain constraints 
on possible imperfections 
in the source and the detectors used in the protocol by the 
two legitimate users 
(Alice and Bob). One way to deal with such imperfections is 
to absorb their effect 
into the attack employed by a potential eavesdropper (Eve). 
In this spirit, most of the 
security proofs assume that any flaws due to imperfections in 
the source and/or the 
detectors do not depend on the bases used in the protocol i.e, 
they do not reveal any information about the basis-choice to Eve. 
Unfortunately, such security proofs are not directly applicable 
to practical implementations of the protocols as typical 
imperfections can be basis-dependent \cite{note1}.

In particular, due to the lack of efficient single-photon sources, 
most of the current realizations of QKD 
protocols use as information carriers weak coherent pulses (WCPs), 
with a sufficiently low probability of containing more than one 
photon \cite{RMP}. Multiphoton pulses, however, threaten the security 
of the  QKD protocols as they can be exploited cleverly by Eve to gain perfect 
information about part of the exchanged random key without 
being detected \cite{L00,BLMS00}. 
To this end, she may launch the so-called photon-number-splitting 
(PNS) attack which, after the announcement of the bases used during 
preparation, enables her to obtain full information about 
the bit encoded in each of the multiphoton pulses \cite{L00,BLMS00}.  
In that respect, each multiphoton signal can be viewed as a 
tagged signal (qubit) which will yield its complete 
information to Eve without introducing detectable errors in the sifted key.
Finally, even today's available single-photon detectors are not 
ideal \cite{RMP}. At telecommunication wavelengths, for example, detection 
efficiencies are typically much smaller than unity while high dark-count 
rates severely limit the maximum distances over which a secret random 
key can be distributed by means of optical fibers 
\cite{RMP,L00,BLMS00,FGSZ01}.

In an effort to bring unconditional security proofs closer to 
practical QKD implementations, 
recent proofs relax the assumption about basis-independent 
eavesdropping \cite{GLLP00,ILM01}. In this context, 
Gottesman, Lo, L\"utkenhaus, and Preskill (GLLP) derived 
a general expression for the asymptotically achievable 
secret-key generation rate for 
the four-state protocol, under the assumption of weakly basis-dependent 
eavesdropping attacks \cite{GLLP00}. Among many types of imperfections,  
the GLLP unconditional security proof takes into account possible tagging 
at Alice's source. From the technical point of view, the GLLP investigation 
concentrates on CSS-based post-processing i.e., error correction and 
privacy amplification protocols involving one-way classical communication 
only and whose achievable secret-key rates result from random encoding 
and decoding by asymmetric Calderbank-Shor-Steane (CSS) quantum 
codes \cite{CSS}. In fact the security is first established in the 
framework of an associated protocol based on a CSS-like 
one-way entanglement purification protocol 
(see definition 4 of Ref. \cite{LoGott}), which is mathematically 
equivalent to CSS quantum codes \cite{BDSW}. Subsequently, the 
entanglement-based 
protocol is reduced to the standard four-state prepare-and-measure scheme 
without compromising security.  

Motivated by these results, in this paper we investigate to which extent 
maximum achievable 
distances of secret-key distribution in the presence of imperfections,  
can be increased  
by additional use of an error-rejection procedure involving two-way 
classical communication. 
For this purpose we concentrate on the aforementioned types of 
experimentally relevant imperfections namely, 
tagging of qubits at Alice's source, dark counts, low efficiency of 
Bob's detectors and losses in the quantum channel connecting them. 
As a particular example 
of an error-rejection procedure we adopt the so-called B-steps of 
the recently proposed two-way post-processing protocol of 
Gottesman and Lo \cite{LoGott}. 
In our subsequent investigation we discuss the four-state, 
the six-state, and the decoy-state \cite{Wang,MaQi05} 
QKD protocols. As a main result it will be demonstrated 
that with the help of a 
succession of B-steps followed by a CSS-based post-processing, the 
achievable distances of secret-key 
generation in optical fibers can be enhanced significantly. 

This paper is organized as follows: 
In Section \ref{Sec2}, we briefly recapitulate basic facts about 
practical QKD implementations and highlight the main (disastrous) 
effect of dark counts on the rates for secret-key generation.
In Section \ref{Sec3} we discuss the 
quantum state of Alice and Bob after a powerful eavesdropping attack 
which takes into account tagged signals, losses and dark counts 
at Bob's detection unit.
In Section \ref{Sec4} the influence of B-steps onto this quantum 
state and its resulting secret-key generation rate 
is investigated. For this latter purpose we focus on a post-processing 
protocol combining B-steps and asymmetric CSS codes. 
The degree to which such a post-processing can suppress the disastrous 
effect of dark counts on the rates for secret-key generation is 
investigated numerically. 

\section{Practical QKD implementations}
\label{Sec2}
In this section, for the sake of completeness, we briefly summarize 
basic facts about 
practical QKD, which are essential for the subsequent discussion. 
In particular, we establish a model for possible imperfections in 
practical implementations of the four- and the six-state QKD protocols,  
and discuss asymptotic secret-key generation rates. 

\subsection{Ideal QKD protocols} 
\label{Sec2.1}
Let us start with a summary of the ideal prepare-and-measure 
four-state and six-state QKD protocols, which typically involve 
three stages. In the 
{\em distribution stage}, Alice encodes her random bit-string in a 
random sequence of non-orthogonal signal states (e.g., polarized 
single photons). Such a preparation involves two mutually unbiased 
bases (MUBs) in the four-state protocol and three in the fully 
symmetric six-state protocol. A first {\em raw key} is established 
when Bob measures each received signal at random in one of the 
possible bases and registers his outcomes. 
In the {\em sifting stage}, Alice and Bob reject all 
(ideally half for BB84 and 2/3 for the six-state protocol) 
bits originated from measurements in bases different from the 
preparation ones. Finally, Alice and Bob post-process this 
{\em sifted key} to distill a secret key. The {\em post-processing stage} 
typically involves error-correction and privacy amplification.
 
Following \cite{GLLP00,LoGott}, throughout this work 
we adopt the {\em equivalent} en\-ta\-ngle\-ment-based versions  
of the prepare-and-measure schemes \cite{BBM,Nielsen}, based 
on a two-way CSS-like entanglement purification protocol (EPP). 
Let us start by recapitulating briefly the main steps involved in them. 
Alice prepares $N$ qubit-pairs in the Bell state 
$\ket{{\rm \Phi^{+}}}^{\otimes N}$ \cite{bell}, where 
$\ket{\rm\Phi^{+}} = (|0\rangle_{\rm A}\otimes |0\rangle_{\rm B} + 
|1\rangle_{\rm A}\otimes |1\rangle_{\rm B})/\sqrt{2}$ 
is a  simultaneous eigenstate of the two Pauli operators 
${\cal X}_{\rm A}\otimes{\cal X}_{\rm B}$ and 
${\cal Z}_{\rm A}\otimes{\cal Z}_{\rm B}$, where 
${\cal Z}=\ket{0}\bra{0}-\ket{1}\bra{1}$ and 
${\cal X}=\ket{0}\bra{1}+\ket{1}\bra{0}$.
She keeps half of each pair (denoted by A) and sends the other 
half (denoted by B) to Bob in one of 
the, say $\beta$, possible MUBs. In other words she applies 
a random rotation ${\cal R}_{\rm B}^{b}$, where 
the random variable $b\in\{0,\ldots,\beta\}$ \cite{LoGott}. 
From now on, we refer to the eigenstates of ${\cal Z}$, 
i.e. $\{\ket{0},\ket{1}\}$, as the Z-basis (computational basis). 
As is well known, the two MUBs $(\beta=2)$ involved in the 
four-state protocol \cite{BB84} 
are related via the Hadamard transformation 
${\cal H}=\sum_{i,j}(-1)^{ij}\ket{i}\bra{j}/\sqrt{2}$ 
where $i,j\in\{0,1\}$ \cite{LoGott,NAG06}, 
while the three MUBs $(\beta=3)$ of the six-state protocol 
\cite{Bruss6} are related via 
successive actions of the unitary operator 
${\cal T} = \sum_{j}\left [
\ket{j}\bra{0}-{\rm i}(-1)^{j}\ket{j}\bra{1}\right ]/\sqrt{2}$ 
\cite{LoGott,NAG06}.  
Hence, in the former case ${\cal R}_{\rm B}={\cal H}_{\rm B}$ 
whereas in the latter ${\cal R}_{\rm B}={\cal T}_{\rm B}$.  

After Bob has received all the transmitted qubits, Alice announces 
the sequence of rotations she performed and Bob undoes all of them. 
Subsequently Alice and Bob randomly permute their qubit-pairs 
so that their resulting $N$-pair quantum state becomes 
{\em permutation invariant}. 
They select a random subset of their pairs and they measure 
each one of them along the Z-basis, to estimate the qubit 
error probability $\delta$ of the residual qubit-pairs \cite{NAG06}.  
If $\delta$ exceeds a certain threshold, secret-key 
distillation cannot be guaranteed and the protocol is aborted. 
Otherwise, Alice and Bob perform a two-way CSS-like EPP to extract pure 
(high-fidelity) entangled pairs, which they measure along the Z-basis 
to obtain the final secret key. 
Most importantly, if the applied two-way CSS-like EPP (such as the one 
considered in the subsequent discussion)  
fulfills the requirements of Theorem 6 in Ref. \cite{LoGott}, the 
entanglement-based protocol can be reduced to a prepare-and-measure 
scheme without compromising the security.  The results we are going 
to present therefore also apply to the corresponding prepare-and-measure 
schemes.

\subsection{A model for imperfections and losses}
\label{Sec2.2}
Typical QKD implementations deviate from the ideal protocols mainly 
in two 
respects: the signal sources are not ideal and the link 
(channel and detectors) between the two legitimate users is lossy 
and noisy. The model we adopt throughout this work 
for the description of such imperfections has been  discussed  thoroughly
in the literature \cite{RMP,L00,BLMS00,FGSZ01}. Here, for the sake of completeness, 
we briefly summarize its main ingredients.

We consider an imperfect source which with probability $p_{\rm tag}$ 
produces tagged qubits (signals), 
in the sense that Eve is capable of extracting from these qubits the information 
which random rotation (basis) has been used by Alice on them before their submission 
to Bob. Thus, Eve is able to measure each one of these qubits in such a way that 
she can unambiguously determine its quantum state without disturbing it i.e., 
without introducing any detectable errors. 
On the contrary, the remaining untagged (ideal) qubits which are produced 
by our source with probability $1-p_{\rm tag}$, do not reveal 
any information to Eve and any intervention of her 
(consistent with quantum mechanics) affecting them will eventually introduce errors. 
Hence, the overall bit-error rate estimated by Alice and Bob 
during the verification test \cite{note2} is basically due to 
untagged qubits only i.e., $\delta = (1-p_{\rm tag})\delta_{\rm b,u}$, 
where $\delta_{\rm b,u}$ is the probability with which 
an untagged qubit contributes to the overall bit-error rate. 
Given the symmetry between all the bases used in the QKD 
protocols under consideration, we expect for the corresponding phase-error 
probability $\delta_{\rm p,u}=\delta_{\rm b,u}$ i.e.,  
$\delta_{\rm p,u}=\delta/(1-p_{\rm tag})$.

A practically relevant special case of tagging is the 
signal sources currently used in various realistic setups,  
which produce polarized phase-randomized WCPs \cite{RMP,L00,BLMS00,FGSZ01}. 
In this case, the photon number distribution $p_i$ 
($i=0,1,...$) of each pulse is Poissonian, 
i.e. $p_i = \exp \left( -\mu \right) \mu^{i}/i!$ 
with $\mu$ denoting the mean photon-number in the pulse. 
Alice, therefore, encodes each of her random bits in a polarized 
WCP which is sent to Bob. However, in addition to 
single-photon pulses such a source may produce
multiphoton pulses and in that respect it deviates from 
the ideal single-photon source. 
More precisely, single-photons are produced with probability 
$p_1$ while multiphoton pulses with probability $p_{\rm tag}=1-p_0-p_1$.
As will be explained in detail later on, Eve can obtain 
full information on all the bits encoded in multiphoton pulses 
by means of the so-called PNS attack. For a source producing 
WCPs therefore, multiphoton and single-photon pulses can be 
viewed as tagged and untagged qubits, respectively. 
Typically in WCP-based QKD implementations $\mu$ is chosen 
sufficiently small, so that the WCP source imitates an 
ideal single-photon source as close as possible 
\cite{RMP}. The limitations of our model for the source will 
be discussed later on in Section \ref{Sec5}.

In addition to imperfect signal sources, realistic setups 
involve imperfect quantum channels and detectors. 
As a result, the raw-key rate $P_{\exp}$ 
(i.e., the probability for a single detection event to occur at Bob's 
site), is in general distance-dependent and less than unity. 
More precisely, $P_{\exp}$ has contributions from 
both real signals arriving at Bob's detector and dark counts. 
In the adopted model, and for the aforementioned WCP source, 
actual signals trigger Bob's detector with probability 
$P_{\exp }^{\rm signal}=1-\exp \left( -\mu\eta _{\rm c}\eta _{\det }\right)$, 
where $\eta _{\rm c}$ and $\eta _{\det}$ denote the transmission 
efficiency of the relevant quantum channel and the 
detection efficiency of Bob's detector, respectively. 
For QKD implementations at telecommunication wavelengths, 
$\eta _{\det}\sim 0.1-0.2$ and $\mu\ll 1$ while the 
quantum channels are optical fibers for which 
\bea
\eta _{\rm c}=10^{-\left( \alpha l+L_{\rm c}\right) /10}.
\label{eta}
\eea
Thereby, $\alpha $ denotes a polarization independent loss coefficient  of the fiber, 
$l$ is the length of the fiber, and  $L_{\rm c}$ characterizes a distance-independent
loss of the channel. Moreover, the total dark-count probability for Bob's detection unit 
involving two identical detectors is $P_{\exp }^{\rm dark}\sim 10^{-4}-10^{-5}$. 
Hence, we typically have \cite{RMP,L00,BLMS00,FGSZ01}
\bea 
P_{\exp } \approx P_{\exp }^{\rm signal}+P_{\exp }^{\rm dark}=
1-{\rm e}^{-\mu\eta _{\rm c}\eta _{\det }}+P_{\exp }^{\rm dark}. 
\label{Pexp_apr}
\eea
Clearly, for an ideal link involving a lossless quantum channel and 
ideal detectors we have $P_{\exp}=1-{\rm e}^{-\mu}$. 

The overall bit-error rate in the sifted key has also two 
contributions and is modeled by \cite{RMP,L00,BLMS00,FGSZ01} 
\begin{equation}
\delta =\delta_{\rm opt}+\delta_{\rm det}=\frac{\delta_0~P_{\exp }^{\rm signal}
+\frac{1}{2}P_{\exp }^{\rm dark}}{P_{\exp }}. 
\label{errorrate}
\end{equation}
The first contribution is independent of the 
transmission distance and is a measure of the optical quality of the whole setup. 
In particular, the constant $\delta_0$ accounts for possible alignment errors, 
polarization diffusion or fringe visibility. 
The second contribution $\delta_{\rm det}$, originates from dark counts at 
Bob's detectors, with the factor $1/2$ indicating that a dark count 
represents one of the two possible random measurement results of Bob. 
Hence, an error will be generated in half of the cases only.
In the most pessimistic scenario usually adopted in security proofs,  
all the error rate $\delta$ is attributed to Eve. 

Finally, any imperfections, losses, and noise significantly 
affect the fraction of tagged qubits arriving at Bob's site. 
In general, the new (effective) tagging probability $\Delta$, 
can be expressed in terms of the parameters 
characterizing the channel, the source and the detectors. 
An upper bound on $\Delta$, for example,  may be obtained 
by the following consideration, in the case of a photon source emitting 
phase-averaged WCPs \cite{L00,BLMS00}. An eavesdropper, Eve, 
with unlimited power  
may not only obtain perfect information about all the classical 
bits originating from multiphoton pulses but  
she may also increase the fraction of these multiphoton pulses as 
much as possible without affecting Bob's expected click-rate probability. 
For this purpose she can replace the lossy quantum channel
by a perfect one (i.e., $\eta_{\rm c}=1$) so that all multiphoton pulses are 
transmitted perfectly. In order to keep $P_{\exp}$ constant she has to block an 
appropriate number of single-photon pulses. Thus, the maximum probability of 
tagged pulses arriving at Bob's detector, which Eve can have perfect knowledge 
about, is given by \cite{L00,BLMS00}
\begin{equation}
\Delta \approx\frac{1-\left( 1+\mu\right) \exp \left( -\mu\right) }{P_{\exp }}, 
\label{DeltaG}
\end{equation}
while the corresponding probability for single-photon pulses is given by $(1-\Delta)$, 
so that they sum up to unity.   

\subsection{Asymptotic secret-key generation rates} 
\label{Sec2.3} 
As shown by GLLP in Ref. \cite{GLLP00}, losses and weak basis-dependent 
imperfections such as tagging do not render any of the QKD protocols 
under consideration insecure, but they affect the 
rates for secret-key generation. 
More precisely, for one-way CSS-based post-processing it has been shown that 
a secret key can be generated by Alice and Bob with the asymptotic rate 
\begin{eqnarray}
R_{\rm CSS} &=& \frac{P_{\exp}}{\beta}\left[ 1-\Delta -H(\delta)-\left(
1-\Delta \right) H\left(\delta_{\rm p, u}\right) \right],
\label{GLLPRate}
\end{eqnarray}%
where $H(x) := -x\log_2 x - (1 - x)\log_2 (1 - x)$ is the 
binary Shannon entropy. 

Let us briefly analyze the quantities entering this expression. 
First of all, in the most pessimistic scenario, all the errors detected by Alice and Bob 
during the verification test are due to Eve's intervention. Furthermore, 
assuming that Eve has maximized the contribution of the tagged qubits in the 
sifted key by replacing the lossy channel with a perfect one,  
the estimated bit-error rate appearing in (\ref{GLLPRate}) is now given by 
$\delta = (1-\Delta)\delta_{\rm b,u}$, while the phase-error probability 
for an untagged qubit reads   
\bea
\delta_{\rm p,u}=\delta/(1-\Delta),
\label{dpu}
\eea 
accordingly. 
The raw-key rate is given by (\ref{Pexp_apr}), while the factor $1/\beta$ accounts 
for the fraction of the raw bits 
passing the sifting procedure in a typical prepare-and-measure scheme.  
Clearly, for the four-state (two-basis) QKD protocol we have $\beta=2$, whereas 
for the six-state (three-basis) protocol $\beta=3$. 
Moreover, it has to be noted that for post-processing procedures taking into 
account possible correlations between bit-flip and phase error, the 
rate (\ref{GLLPRate}) can be improved \cite{L01}. 
However, throughout this work we adopt for both protocols the worst-case 
scenario which corresponds to having no such correlations thus implying 
zero mutual information between bit-flip and phase errors. 
 
\begin{figure}
\vspace{0.7cm}\hspace{1.0cm}\resizebox{0.75\columnwidth}{!}{%
  \includegraphics{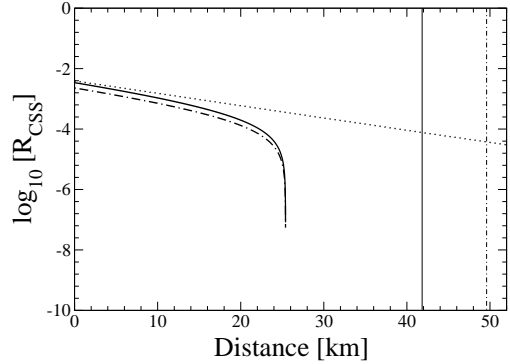}}
\caption{ Achievable secret-key rates as given by 
equation $(\ref{GLLPRate})$, for non-ideal implementations of the 
four-state (full curve) and the six-state (dot-dashed curve) QKD protocols: 
Error correction and privacy amplification are performed by means of 
asymmetric CSS codes which involve one-way classical communication only. 
The vertical lines indicate the maximum allowed distances 
for secret-key generation as determined by equation (\ref{MaxDistance}) 
for the four-state (solid line at $\sim 42~{\rm km}$ ) and 
the six-state protocol (dot-dashed line at $\sim 50~{\rm km}$).
Also shown is the secret-key rate of the four-state protocol
in the absence of dark counts (dotted curve). All relevant parameters are 
chosen as in the experiment of Ref. \cite{KTH} i.e., 
$\alpha  =0.2\text{dB}/{\rm km}$, $L_{\rm c}= 1\text{dB}$, $\delta_0=1\%$, 
$P_{\exp}^{\rm  dark} =2\times 10^{-4}$, $\eta_{\det }=0.18$. At each 
distance the mean number of photons $\mu$ is optimized such that the 
corresponding rate is maximal.} 
\label{OnewayBB84+sixstate}       
\end{figure}

Typical behavior of the secret-key generation rate as a function of the distance 
(i.e., the length of the optical fiber $l$) is depicted in Figure \ref{OnewayBB84+sixstate}. 
Using equations (\ref{eta}-\ref{DeltaG}) and (\ref{dpu}), we plot 
the rate $R_{\rm CSS}$ (as determined by equation \ref{GLLPRate}) for the four- and the 
six-state QKD protocols. 
A sudden drop of the key generation rate at about $25~{\rm km}$ is clearly apparent in 
both protocols. 
A comparison with the corresponding secret-key generation rate of the four-state QKD protocol
in the absence of dark counts (dotted curve) exhibits that this drop originates from dark counts.
Indeed, in the case of a lossy quantum channel, the contribution to $\delta$ due to signals decreases 
with increasing $l$, so that eventually almost all the contributions to the 
error rate $\delta$ originate from dark counts. 
At the critical distance of $25~{\rm km}$ the contribution of dark counts becomes 
dominant and almost all the key is lost by error correction and privacy 
amplification \cite{RMP,L00}. The critical distance turns out to be the same  
for the two protocols as a result of equation (\ref{GLLPRate}) which was used 
for both of them. However, as discussed in \cite{L01}, 
the secret-key rate for the six-state protocol 
can be improved by means of a post-processing procedure 
which takes into account correlations between bit-flip and phase errors.

An upper bound on achievable distances can be obtained from the maximally 
tolerable error rates for single-photon pulses \cite{RMP,L00}.
More precisely, given that Eve has full information on the error-free 
part of the key stemming from multiphoton pulses, Alice and Bob can 
extract a secret key only if they can prove the presence of 
quantum correlations (provable entanglement) in the remaining 
part of the sifted key originating entirely from single-photon 
pulses \cite{CLL,AG}. 
However, this is possible only if the corresponding (rescaled) error rate 
$\delta/(1-\Delta)$ does not exceed $1/4$ for the four-state and 
$1/3$ for the six-state QKD protocol i.e.,  
\bea
\frac{\delta }{\left( 1-\Delta \right) }<\frac{\beta-1}{2\beta},\quad
{\rm for}\quad \beta\in\{2,3\}.
\label{MaxDistance}
\eea
This is a generalization of the necessary conditions for secret-key generation 
in the context of the four- and the six-state protocols respectively, 
in the absence of tagging \cite{NAG06,CLL,AG}. 
Indeed, an intercept-resend eavesdropping 
attack can always break entanglement between Alice and Bob giving rise to 
an error rate  
$\delta\geq (1-\Delta)(\beta-1)/2\beta$.

The necessary condition (\ref{MaxDistance}) limits the distances up to 
which a secret key can be distilled since both $\Delta $ and $\delta $ 
depend on the length of the optical fiber connecting Alice and Bob. 
This bound is indicated in Figure \ref{OnewayBB84+sixstate} by the solid 
vertical line for the
four-state and by the dot-dashed vertical line for the six-state
protocol. In Section \ref{Sec4} we will demonstrate how the gap between these 
borders and the drop of the secret-key generation 
rates due to dark counts can be
decreased considerably by applying a two-way error-rejection procedure 
before switching to one-way CSS-based post-processing. Before that, we have to derive the 
quantum state shared between Alice and Bob at the end of the 
distribution stage.

\section{Formulation of QKD with tagged qubits}
\label{Sec3} 
In order to derive the quantum state of Alice and Bob immediately before 
the post-processing stage under a realistic scenario, we have to take 
into account the influence of a tagging source and possible imperfections  
in the link (quantum channel and detectors). Most importantly, 
we also have to consider in detail Eve's strategy which can, 
in principle, take full advantage of all imperfections and losses.  
Following \cite{GLLP00,LoGott}, we adopt the entanglement-based version 
of the four- and the six-state QKD protocols described in Section 
\ref{Sec2.1}.

\subsection{An optimal eavesdropping strategy}
\label{Sec3.1} 
Consider the tagging scenario described at the end of  
Section \ref{Sec2.2}, in which tagged qubits arrive at Bob's site  
with probability $\Delta$. 
Let also $N$ be the total number of qubit-pairs shared between Alice and Bob 
at the end of the distribution stage. For sufficiently large values of $N$, 
we expect that $N_{\rm u}\approx(1-\Delta)N$ pairs involve untagged qubits 
and $N_{\rm t}\approx \Delta N$ pairs involve tagged qubits 
(to be referred to hereafter as the tagged qubit-pairs).
In general, the form of the reduced state of all $N$ pairs,  
$\rho_{\rm tot}^{(N)}$, depends on the eavesdropping attack employeed by Eve. 

As we discussed earlier, a tagged qubit reveals to Eve its 
basis of preparation i.e., which random rotation has Alice applied on it 
before its submission to Bob. Thus, Eve is able to measure each 
tagged qubit in such a way that she can unambiguously determine 
its quantum state without disturbing it i.e., without introducing any 
errors. On the contrary, the remaining untagged qubits are ideal for 
Alice and Bob, in the sense that they do not reveal 
any information to Eve. In particular, given that each untagged qubit 
is randomly prepared in non-orthogonal states, 
information gain for Eve is only possible at the expense of disturbing 
its state, thus introducing errors in the sifted key \cite{Nielsen}.
It has to be noted here that in the model under consideration, the source simply 
tags any $\Delta N$ of the signals in an uncorrelated and independent way. 
In other words, we do not allow for coherent superpositions of tagging procedures 
or other highly correlated basis-dependent imperfections \cite{GLLP00}. 
Moreover, we assume that apart from the tagging scenario we have just described, 
the source behaves in a perfect way while the tagged signals do not 
convey any information about the untagged ones.  
In this framework, we restrict ourselves to a particular class of 
powerful eavesdropping strategies where Eve treats tagged and 
untagged qubits separately. 
Indeed, tagging allows Eve to attack  a fraction of the qubits 
without introducing errors. Any other eavesdropping 
strategy which is consistent with quantum mechanics  and does not take into 
account the tagging may only result in higher error rates in the sifted key. 
Finally, given that Eve can have full information on all the 
bits encoded in tagged qubits, she may launch the most powerful attack 
i.e., a coherent attack, on the remaining untagged qubits to extract as 
much information as possible about the final key. 

Therefore, as long as Eve attacks the sets of tagged and untagged qubits 
separately, these sets are not entangled between each other.  
From now on, the reduced state of all tagged qubit-pairs is denoted by 
$\rho_{\rm t}^{(N_{\rm t})}$ whereas the corresponding state of all 
untagged qubit-pairs is denoted by $\rho_{\rm u}^{(N_{\rm u})}$. 
Our task is to estimate the precise form of these states 
and to this end we have to consider a particular tagging scenario.
 
\subsubsection{Attack on tagged qubits}
We will focus on the practically relevant special case of tagging 
discussed in Section \ref{Sec2.2} that is, a source which 
produces phase-randomized WCPs. 
For each multiphoton pulse sent by Alice to Bob, first of all Eve can 
measure the photon number.
She can do this by means of a quantum non-demolition measurement 
without introducing any disturbance. In a second step, Eve can extract 
from each of these multiphoton pulses one photon, as described in the 
appendix of Ref. \cite{L00}, which  is stored in a 
quantum memory while the remaining signal 
is sent to Bob. After the announcement of the bases used during 
preparation, Eve measures each of her photons in the correct basis and 
obtains full information about the corresponding encoded bit. 
 
Clearly, by such a PNS attack Eve can eventually determine 
all the key bits which originate from multiphoton pulses 
without introducing any bit-flip errors. Moreover, she may 
adjust her attack so that her intervention 
remains undetected, even if Alice and Bob proceed to monitor the 
complete photon-number distribution \cite{LJ02}. Hence, the PNS attack turns 
out to be Eve's optimal attack on the multiphoton pulses \cite{L00}.

Now we turn to estimate the reduced state of Alice and Bob for all 
tagged qubit-pairs, $\rho_{\rm t}^{(N_{\rm t})}$.
Since Eve attacks each tagged pair individually, we have 
$\rho_{\rm t}^{(N_{\rm t})}=\sigma^{\otimes N_{\rm t}}$. Therefore, it  
is sufficient to consider one of these qubit pairs. After Bob has undone the 
rotation applied by Alice on his qubit, the purified quantum state 
of a tagged qubit-pair for which Bob's half has suffered a 
PNS attack is of the form
\bea
|\chi\rangle_{\rm ABE} &=& 
\frac{1}{\sqrt{2}}\left (|0\rangle_{\rm A}\otimes|0\rangle_{\rm B}\otimes
|\tilde{0}\rangle_{\rm E} 
+ |1\rangle_{\rm A}\otimes|1\rangle_{\rm B}\otimes
|\tilde{1}\rangle_{\rm E}\right )\nonumber\\ 
&\equiv&\frac{1}{\sqrt{2}}\left (|\rm\Phi^+\rangle\otimes|0\rangle_{\rm E} 
+ |\rm\Phi^-\rangle\otimes|1\rangle_{\rm E}\right).
\label{EveEntang} 
\eea 
This state can be obtained, for example, in the context of the Jaynes-Cummings 
Hamiltonian discussed in the appendix of Ref. \cite{L00}.
Thereby, Eve's pure ancilla states
$|0\rangle_{\rm E} = \left (|\tilde{0}\rangle_{\rm E} +|\tilde{1}\rangle_{\rm E}\right )/\sqrt{2}$ and 
$|1\rangle_{\rm E} = \left (|\tilde{0}\rangle_{\rm E} -|\tilde{1}\rangle_{\rm E}\right )/\sqrt{2}$ are orthogonal. 
The Bell state
$|\rm\Phi^-\rangle = (|0\rangle_{\rm A}\otimes |0\rangle_{\rm B} - 
|1\rangle_{\rm A}\otimes |1\rangle_{\rm B})/\sqrt{2}$
characterizes the phase errors introduced by Eve's ideal attack.
The equal amplitudes of magnitude $1/\sqrt{2}$ reflect the fact that Eve does not perturb Alice's and Bob's measurement statistics by her attack.
Correspondingly, the reduced quantum state of Alice and Bob resulting from such an ideal attack is 
a random mixture of the ideal Bell state $|\rm\Phi^+\rangle$ and the corresponding 
phase-flipped Bell state $|\rm\Phi^-\rangle$, i.e.,
\begin{eqnarray}
\sigma &=& \frac{1}{2}(|\rm\Phi^+\rangle\langle \rm\Phi^+| +|\rm\Phi^-\rangle\langle \rm\Phi^-| ).
\label{sigma}
\end{eqnarray}
The separability of the state (\ref{sigma}) reflects the fact that Eve 
has a perfect copy of Bob's qubit and thus secret-key distillation 
is impossible \cite{NAG06,CLL,AG}.

\subsubsection{Attack on untagged qubits}
Being able to obtain full information on the fraction of the key originating 
from tagged qubits, Eve may attack the remaining $N_{\rm u}$ 
untagged qubits coherently 
in order to obtain additional information on the final key.
In general, at the end of such a coherent attack the reduced state of the 
untagged pairs $\rho_{\rm u}^{(N_{\rm u})}$ has a rather complicated form. 
In particular each pair can be  
entangled with Eve's ancilla as well as with other untagged qubit-pairs. 

However, the key point is that Alice and Bob randomly permute all 
their (tagged and untagged) pairs immediately after Bob has announced the reception of all his 
qubits. As a result, the state  $\rho_{\rm u}^{(N_{\rm u})}$ becomes 
permutation invariant. Hence, according to Lemmas 2 and 3 
(including the related proofs) of Ref. \cite{LoGott}, it suffices for our purposes to consider 
uncorrelated Pauli attacks only, where Eve applies a random Pauli operator 
independently on each submitted qubit. In particular, she applies 
${\cal X}$ with probability $q_{\rm x}$, ${\cal Z}$ with probability $q_{\rm z}$, 
${\cal Y}\equiv{\rm i}{\cal XZ}$ with probability $q_{\rm y}$, and the identity ${\cal I}$ 
with probability 
$q_{\rm I}=1 - q_{\rm x} -q_{\rm y} - q_{\rm z}$. The resulting state of 
an uncorrelated Pauli attack is 
therefore of the form 
$\rho_{\rm u}^{(N_{\rm u})}=\tau^{\otimes N_{\rm u}}$, where 
the single-pair state $\tau$ is diagonal in the Bell-basis \cite{bell} i.e., 
\begin{eqnarray}
\tau &=& q_{\rm I}|{\rm\Phi^+}\rangle \langle{\rm \rm\Phi^+|}+ 
q_{\rm z} |{\rm\Phi^-\rangle} \langle {\rm\Phi^-|}\nonumber\\
&+&q_{\rm x} |{\rm\Psi^+\rangle} \langle {\rm\Psi^+|}+ 
q_{\rm y} |{\rm\Psi^-\rangle} \langle {\rm\Psi^-|}.
\label{tau}
\end{eqnarray}

Note now that due to Alice's random rotation the quantum state $\tau$ 
is also rotation invariant i.e., it is symmetrized 
with respect to the corresponding group of unitary transformations 
${\cal G} = 
\{{\cal R}_{\rm A}^b \otimes {\cal R}_{\rm B}^b~|~b=0,\ldots,\beta\}$. 
This symmetry, implies additional constraints on the (error) probabilities 
$(q_{\rm x},q_{\rm y},q_{\rm z})$ of the state (\ref{tau}). 
In particular, for the four-state 
protocol we have $q_{\rm x} = q_{\rm z}$ whereas for the more 
symmetric six-state 
protocol $q_{\rm x} = q_{\rm y}= q_{\rm z}$. Thereby, these 
symmetry constraints are 
characteristic for the two(three) MUBs used in the 
four(six)-state QKD protocol. It is worth noting, however, that 
this rotation-invariance does not apply to the case of PNS attacks, 
as the tagged qubits inform Eve about their bases of preparation. 
Eve can always therefore follow the rotations applied by Alice and 
remain undetected. 

\subsection{Alice's and Bob's point of view}
\label{Sec3.2} 
Let us now assume that Alice and Bob do not have Eve's technology and thus are 
not able to distinguish between tagged and untagged qubit-pairs. 
In other words, from their point of view all the pairs are equivalent. 
They only know that their imperfect source produces ideal qubits with 
some probability, and tags the qubits otherwise. 
Formally speaking, Alice and Bob share $N$ qubit-pairs in the 
quantum state 
\bea\rho_{\rm tot}^{(N)}= \frac{1}\Pi\sum_{\Pi}
\Pi\left (\sigma^{\otimes N_{\rm t}} 
\otimes\tau^{\otimes N_{\rm u}}\right )\Pi^\dag,
\label{rtall}
\eea 
where the summation runs over all possible permutations and expresses 
Alice's and Bob's ignorance about the precise location of the tagged 
pairs within the block of $N$ pairs. 
In the limit of large $N$ we have $N_{\rm u}\approx(1-\Delta)N$ and 
$N_{\rm t}\approx \Delta N$, thus obtaining from equation (\ref{rtall})   
\bea
\rho_{\rm tot}^{(N)}\approx\rho^{\otimes N},
\label{ansatz}
\eea
where 
\bea
\rho &=& \Delta\sigma + (1 - \Delta)\tau.
\label{ansatz2}
\eea
An easy way to see this, is by using the binomial theorem since 
the density operators $\sigma$ and $\tau$ commute. 

Hence, in view of the state (\ref{ansatz}-\ref{ansatz2}), 
the overall bit-error 
rate estimated by Alice and Bob 
during the verification test by random pair-sampling and 
measurements along the Z-basis \cite{note2} is given by
\begin{eqnarray}
\delta = (1-\Delta)\delta_{\rm b,u} = (1-\Delta)(q_{\rm x} + q_{\rm y}),
\label{qber2}
\end{eqnarray}
where $\delta_{\rm b,u}$ is the error probability for a single 
untagged pair as determined by its state (\ref{tau}). 
In view of the symmetry between all the bases used in the protocols 
we also have for the phase-error probability of the untagged pairs 
$\delta_{\rm p,u}=\delta_{\rm b,u}$.

Having derived the state shared between Alice and Bob at the beginning 
of the post-processing stage, we now turn to discuss 
asymptotic secret-key generation rates in the context of a two-way 
CSS-like EPP.

\section{Increasing secure distances using two-way post-processing}
\label{Sec4}
In this section, we demonstrate how one can suppress the disastrous effect of  
dark counts (exhibited as a sudden drop of $R_{\rm CSS}$ in 
Figure \ref{OnewayBB84+sixstate}), thus 
increasing the 
distance over which a secret key can be distributed. 
Our approach relies on a two-way error-rejection procedure followed by a 
one-way 
CSS-like EPP. 

\subsection{Error-rejection with two-way classical communication}
\label{Sec4.1}
The error-rejection procedure under consideration is the so-called B-step 
entering a two-way post-processing of the Gottesman-Lo-type \cite{LoGott,C02}. 
It is basically a purification process with 
two-way classical communication and its properties have been 
thoroughly 
discussed in the literature \cite{LoGott,C02,DEJ,BDSW2,Kedar}. In all these 
investigations, the authors mainly focus on the influence of the B-steps on 
the error rates as all the involved qubit-pairs are identical. 
In our case, however, the situation is substantially different as the 
qubit-pairs 
involved in a B-step may be tagged or untagged. Given that Eve has full 
information on Bob's 
qubit in the former case, in addition to error rates we have to 
keep track of any changes in the rate of tagged qubit-pairs during B-steps. 

Let us start by briefly recapitulating the stages of a B-step. 
Alice and Bob randomly form tetrads of qubits by pairing up their qubit-pairs.  
Then, within each tetrad they apply a bilateral exclusive-OR operation (BXOR) 
i.e., they apply the local unitary operation 
${\rm XOR}_{{\rm a}\to{\rm b}}: \ket{x}_{\rm a}\otimes 
\ket{y}_{\rm b}\mapsto\ket{x}_{\rm a}\otimes 
\ket{x\oplus y}_{\rm b}$,
on their halves. Thereby, $\oplus$ denotes addition modulo $2$ while 
${\rm a}$ and ${\rm b}$ denote the control and target qubit, respectively. 
Accordingly, for the two qubit-pairs constituting the random tetrad 
we have the following map in the Bell basis
\bea
{\rm BXOR}_{{\rm a}\to{\rm b}}:\ket{{\rm\Psi}_{i,j}^{\rm (a)}}\otimes\ket{{\rm\Psi}_{x,y}^{\rm (b)}}\mapsto
\ket{{\rm\Psi}_{i,j\oplus y}^{\rm(a)}}\otimes\ket{{\rm\Psi}_{i\oplus x, y}^{\rm(b)}},
\label{bxor}
\eea
where $i,j,x,y\in\{0,1\}$ and the Bell states are denoted by 
$\ket{\rm\Psi_{0,0}}\equiv\ket{\rm\Phi^+}$,
$\ket{\rm\Psi_{0,1}}\equiv\ket{\rm\Phi^-}$,
$\ket{\rm\Psi_{1,0}}\equiv\ket{\rm\Psi^+}$, and 
$\ket{\rm\Psi_{1,1}}\equiv\ket{\rm\Psi^-}$.  
Subsequently, Alice and Bob measure their target qubits (b)  in the Z-basis  
and compare their outcomes. The target pair is always discarded while 
the control qubit-pair is kept if and only if their outcomes agree i.e., 
if and only if $i=x$. In general, this procedure is repeated many times 
(many rounds of B-step).

Consider now that Alice and Bob apply the B-step procedure 
we have just described on their pairs before switching to 
a CSS-like EPP. 
Recall also that the quantum state of all 
$N$ qubit-pairs shared between Alice and Bob at this stage of the 
QKD protocol 
has the general tensor-product form given in equations 
(\ref{ansatz}-\ref{ansatz2}). Depending on whether the qubit-pairs
forming a random tetrad are untagged or tagged or only one of them is 
tagged we may distinguish four different cases.
\begin{enumerate}
\item {\em Untagged target and control pairs.} According to equations 
(\ref{ansatz}-\ref{ansatz2}), such a pairing occurs with probability 
$(1-\Delta)^2$, while each of the qubit-pairs is in 
the Bell-diagonal quantum state (\ref{tau}).
Hence, provided that Alice's and Bob's measurements agree, the control pair is kept 
and is mapped again onto a Bell-diagonal quantum state of the same form, 
but with 
renormalized parameters \cite{LoGott}
\begin{eqnarray} 
q^{\prime}_{\rm I} &=&\frac{(q_{\rm I} + q_{\rm z})^2 + (q_{\rm I} - q_{\rm z})^2}{2 Q_{\rm u,s}},\nonumber\\
q^{\prime}_{\rm z} &=&\frac{(q_{\rm I} + q_{\rm z})^2 - (q_{\rm I} - q_{\rm z})^2}{2 Q_{\rm u,s}},\nonumber\\
q^{\prime}_{\rm x} &=&\frac{(q_{\rm x} + q_{\rm y})^2 + (q_{\rm x} - q_{\rm y})^2}{2 Q_{\rm u,s}},\nonumber\\
q^{\prime}_{\rm y} &=&\frac{(q_{\rm x} + q_{\rm y})^2 - (q_{\rm x} - q_{\rm y})^2}{2 Q_{\rm u,s}},
\label{map}
\end{eqnarray} 
where $Q_{\rm u,s} = (q_{\rm I} + q_{\rm z})^2 + (q_{\rm x} + q_{\rm y})^2$ is the probability with 
which the control qubit-pair is kept. Moreover, conservation of probability requires 
the relation $q_{\rm I} + q_{\rm z} + q_{\rm x} + q_{\rm y} = q^{\prime}_{\rm I} + q^{\prime}_{\rm z} + q^{\prime}_{\rm x} + q^{\prime}_{\rm y} =1$. 
\item {\em Tagged target and control pairs.} 
In view of equations (\ref{ansatz}-\ref{ansatz2}) such a pairing takes place with probability 
$\Delta^2$. The two pairs are in the same Bell-diagonal state given by equation 
(\ref{sigma}), and thus the map (\ref{map}) applies also in this case. 
Setting $q_{\rm x}=q_{\rm y}=0$ and $q_{\rm I}=q_{\rm z}=1/2$, we have that  
the control pair always survives and is again tagged i.e., its state 
is given by (\ref{sigma}).
\item {\em Tagged target pair and untagged control pair.} Such a pairing   
occurs with probability $\Delta(1-\Delta)$.
Using the map (\ref{bxor}) and the form of the states $\tau$ and 
$\sigma$ given by equations (\ref{tau}) and (\ref{sigma}) respectively, 
one immediately obtains that for the case under consideration the control 
pair survives with probability $Q_{\rm t,s}=(q_{\rm I} + q_{\rm z})$ and 
is left in a quantum state of the form (\ref{sigma}). 
Knowing that one of the purifications of such a state is 
equation (\ref{EveEntang}), and giving all the purification to Eve 
\cite{Nielsen}, we may conclude that  the state of the surviving control pair  
refers to the tagged state of equation (\ref{EveEntang}). In other words, 
the initially untagged control pair becomes tagged when paired with 
a tagged target pair. 
This is equivalent to the XOR operation of an unknown classical bit $S$ 
with a totally known classical bit $M$. Since the target bit $T=S \oplus M$ is 
announced publically, $S$ becomes perfectly known to Eve. 
\item {\em Untagged target pair and tagged control pair.} This is equivalent to 
the previous case.
\end{enumerate}
In summary, only cases in which both pairs involved in a random 
tetrad are untagged can lead to an untagged surviving qubit-pair which may later 
on result to a secret bit for Alice and Bob. In all other cases, Eve has 
a perfect copy of Bob's surviving tagged qubit. 

A qubit-pair initially prepared in the mixed quantum state (\ref{ansatz2}) 
with $\sigma$ and $\tau$ given by equations (\ref{sigma}) and (\ref{tau}) 
respectively, survives the first B-step with probability  
\begin{eqnarray}
P_{\rm s}^{\prime} &=&\
(1-\Delta)^2 Q_{\rm u,s} + 
2 \Delta (1 - \Delta)Q_{\rm t,s} + \Delta^2.
\label{Pprime}
\end{eqnarray}
Moreover, its new quantum state is given by 
\begin{eqnarray}
\rho^{\prime} &=& \Delta^{\prime} \sigma + (1 - \Delta^{\prime})\tau^{\prime},
\end{eqnarray}
with the renormalized tagging probability 
\begin{eqnarray}
\Delta^{\prime} &=& \frac{[\Delta^2  + 2\Delta (1 - \Delta)(q_{\rm I} + q_{\rm z})]}{P_{\rm s}^{\prime}},
\label{Deltaprime}
\end{eqnarray}
and with the untagged renormalized quantum state
\begin{eqnarray}
\tau^{\prime} &=& 
q^{\prime}_{\rm I}|{\rm \Phi^+\rangle} \langle {\rm\Phi^+|}+ 
q^{\prime}_{\rm z} |{\rm \Phi^-\rangle} \langle \rm \Phi^-|
\nonumber\\
&+&q^{\prime}_{\rm x} |{\rm \Psi^+\rangle} \langle {\rm\Psi^+|}
+q^{\prime}_{\rm y} |{\rm \Psi^-\rangle}\langle {\rm\Psi^-|}
\label{tauprime}
\end{eqnarray}
where the new probabilities $(q_{\rm I}^\prime, q_{\rm z}^\prime, q_{\rm y}^\prime, q_{\rm z}^\prime)$ 
are determined by equations (\ref{map}).
Correspondingly, the bit-error probability of this new quantum 
state is given by
\begin{eqnarray}
\delta^{\prime} = (1 - \Delta^{\prime}) \delta_{\rm b,u}^{\prime} = 
(1 - \Delta^{\prime}) (q_{\rm x}^{\prime} + q_{\rm y}^{\prime} ). 
\label{Qprime}
\end{eqnarray}
As a result of the B-step, however, the probabilities of bit and phase errors 
for an untagged qubit are not equal anymore. 
In particular, we have 
\bea
\delta_{\rm p,u}^{\prime} = (q_{\rm z}^{\prime} + q_{\rm y}^{\prime} ).
\label{QPprime}
\eea

Consider now that immediately after one such B-step Alice and Bob switch 
to a one-way CSS-like EPP to distill a secret key. 
The overall asymptotically achievable secret-key generation 
rate is given by the corresponding modification of equation (\ref{GLLPRate}) i.e.,
\begin{eqnarray}
R_{\rm{BCSS}} &=& \frac{P_{\rm exp}P_{\rm s}^{\prime}}{2\beta}
\left(1 - \Delta^{\prime} - H(\delta^{\prime}) - (1 - \Delta^{\prime})H(\delta_{\rm p,u}^{\prime})\right),
\nonumber\\
\label{BCSS}
\end{eqnarray}
where $\Delta^{\prime}$, $\delta^{\prime}$ and  $\delta_{\rm p,u}^{\prime}$ are given by 
equations (\ref{Pprime}-\ref{QPprime}). The additional factor of $1/2$ accounts for 
the target qubit-pairs which are always thrown away during the B-step.  
With the help of the recursion relations (\ref{map}) and  (\ref{Deltaprime}) 
asymptotically achievable secret-key 
generation rates can also be determined for cases in which B-steps 
are applied iteratively before the final use of the 
one-way CSS-like EPP. In that case, however, the factor of 
$1/2$ should be replaced by $1/2^n$, for $n$ B-steps. 
The rate $R_{\rm{BCSS}}$ is therefore a generalization of the 
GLLP rate $R_{\rm{CSS}}$ to a post-processing where 
the one-way CSS-like EPP is initialized by a number of B-steps. 
Indeed, the rate (\ref{BCSS}) directly reduces 
to the rate (\ref{GLLPRate}) in the absence of B-steps i.e., by 
setting $(q_{\rm I}^{\prime},q_{\rm x}^{\prime},q_{\rm y}^{\prime},q_{\rm z}^{\prime})=(q_{\rm I},q_{\rm x},q_{\rm y},q_{\rm z})$, 
$P_{\rm s}^{\prime}=1$, $\Delta^{\prime}=\Delta$, and dropping the 
factor $1/2$. 

\subsection{Numerical simulations and discussion}
In our simulations, we adopt the most pessimistic approach i.e, we consider 
an eavesdropper 
with unlimited technological power \cite{L00}. 
In particular, we attribute all the estimated 
bit-error rate $\delta$ to Eve, assuming that she possesses the 
corresponding information on the key. We thus give Eve all the power to replace the lossy channel by a 
perfect one (as described in Sect. \ref{Sec2.2}), and to adjust the two contributions 
in $\delta$ (that is, $\delta_{\rm opt}$ and $\delta_{\rm det}$) at her own benefit 
(see also related discussion in Ref. \cite{FGSZ01}).  Formally speaking, combining 
equations (\ref{errorrate}) and (\ref{qber2}), at the beginning of the first B-step we have  
\bea
\delta = (1-\Delta)(q_{\rm x}+q_{\rm y}) = 
\frac{\delta_0~P_{\exp }^{\rm signal}
+\frac{1}{2}P_{\exp }^{\rm dark}}{P_{\exp }}, 
\label{inc1}
\eea
where $P_{\exp}$, $P_{\exp }^{\rm signal}$ and $\Delta$ are defined in 
Section \ref{Sec2.2}. 
However, as we discussed in Section \ref{Sec3.1}, 
the probabilities $(q_{\rm I},q_{\rm x},q_{\rm y},q_{\rm z})$ 
entering the map (\ref{map}) 
are not independent. 
The normalization condition for the state (\ref{tau}) implies that 
\bea
q_{\rm I} = 1 - q_{\rm x} - q_{\rm y} - q_{\rm z},
\label{inc2}
\eea
while due to symmetry between all the bases used in the QKD protocols 
under consideration we have one additional constraint. That is,   
\bea
q_{\rm x}=q_{\rm z}=q_{\rm y} 
\label{inc3}
\eea 
for the six-state protocol, and 
\bea 
q_{\rm x}=q_{\rm z}
\label{inc4}
\eea
for the four-state protocol, respectively. 

In the case of the six-state protocol the constraints 
(\ref{inc1} - \ref{inc3}) fully determine the initial 
values of the probabilities $(q_{\rm I},q_{\rm x},q_{\rm y},q_{\rm z})$. 
More precisely, we have 
\bea
q_{\rm x} &=& q_{\rm y}=q_{\rm z}=\frac{\delta_0~P_{\exp }^{\rm signal}
+\frac{1}{2}P_{\exp }^{\rm dark}}{2(1-\Delta)P_{\exp }},\nonumber\\
q_{\rm I} &=& 1-\frac{3\left (\delta_0~P_{\exp }^{\rm signal}
+\frac{1}{2}P_{\exp }^{\rm dark}\right )}{2(1-\Delta)P_{\exp }}.
\label{inc-six}
\eea
On the contrary, such a unique choice is not possible for the four-state protocol 
and we have one open parameter left that is, $0\leq q_{\rm y}\leq 1$. 
It is known, however, that for the map (\ref{map}), the choice 
$q_{\rm y} = 0$ gives rise to the largest resulting value of 
the phase-error probability  and  to the smallest resulting secret-key 
rate \cite{LoGott}. 
Therefore, in the case of the four-state QKD protocol we can restrict our 
subsequent discussion to the initial condition
\bea
q_{\rm y} &=& 0,\nonumber\\
q_{\rm x} &=& q_{\rm z}=\frac{\delta_0~P_{\exp }^{\rm signal}
+\frac{1}{2}P_{\exp }^{\rm dark}}{(1-\Delta)P_{\exp }},\nonumber\\
q_{\rm I} &=& 1-\frac{2\left (\delta_0~P_{\exp }^{\rm signal}
+\frac{1}{2}P_{\exp }^{\rm dark}\right )}{(1-\Delta)P_{\exp }}.
\label{inc-four}
\eea
Clearly, in both cases  
all the probabilities are distance-dependent. 
Indeed, the larger the distance between Alice and Bob becomes, 
the more power Eve has as she may take full advantage of all 
losses, noise, and imperfections.

\begin{figure}
\vspace{0.7cm}\hspace{1.0cm}\resizebox{0.75\columnwidth}{!}{%
  \includegraphics{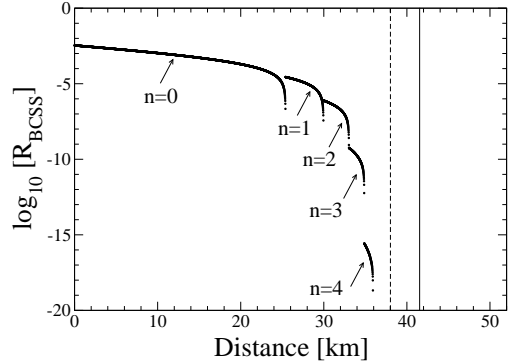}}
\caption{Four-state protocol. Secret-key generation rates resulting 
from multiple 
applications of B-steps followed by one-way CSS-based post-processing: 
The solid vertical line indicates the maximum allowed distances according to
inequality (\ref{MaxDistance}). The dotted line is the asymptotically
achievable distance according to inequality (\ref{MaxDistEPP2}). 
The parameters are the same as in Figure \ref{OnewayBB84+sixstate}.}
\label{BB84KTH1}
\end{figure}

\begin{figure}
\vspace{0.7cm}\hspace{1.0cm}\resizebox{0.75\columnwidth}{!}{%
  \includegraphics{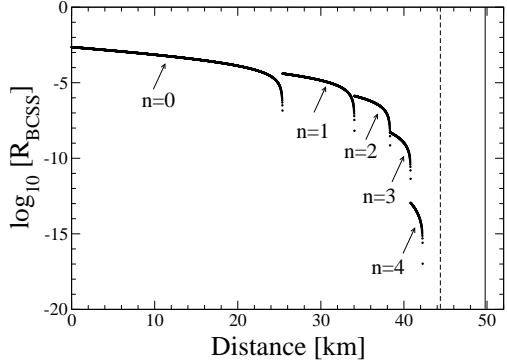}}
\caption{Six-state protocol. The parameters are the same as 
in Figure \ref{BB84KTH1}.}
\label{SixStateKTH1}
\end{figure}

\begin{figure}
\vspace{0.7cm}\hspace{1.0cm}\resizebox{0.75\columnwidth}{!}{%
  \includegraphics{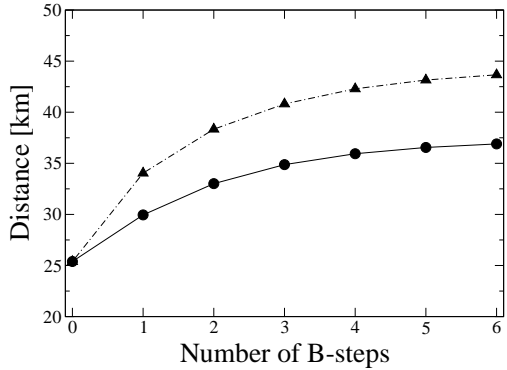}}
\caption{Maximum achievable distance for different numbers of B-steps
for the four-state (lower curve) and the six-state protocol (upper curve).}
\label{nvsdistbbsix}
\end{figure}

We turn now to present and discuss numerical results regarding the effect 
of applied B-steps on the the secret-key rates, for various QKD qubit-based 
protocols. 
For short distances (i.e., length of the fiber $l$) where no B-steps  
are necessary, the secret-key rate  is basically determined by 
equations (\ref{GLLPRate}), (\ref{dpu}) and (\ref{errorrate}), 
as in Section \ref{Sec2.3}. 
However, for secret-key distribution over larger distances application 
of B-steps prior to post-processing by one-way CSS-like EPP is a necessity 
and the corresponding secret-key 
rate  is given by (\ref{BCSS}) combined 
with equations (\ref{Pprime}-\ref{QPprime}) and the initial  
condition for B-steps (\ref{inc-six}) or (\ref{inc-four}). 
In any case, for a given distance we optimize the mean number of photons 
to obtain the maximum possible secret-key rate.

The influence of different numbers of B-steps on $R_{\rm BCSS}$ is depicted in 
Figures \ref{BB84KTH1} and \ref{SixStateKTH1} for the four- and 
the six-state protocol, respectively.
Whereas for the parameters chosen 
in the absence of any B-steps the maximum possible
distance over which a secret key can be distilled 
with a significant rate is of the order of $25~{\rm km}$ for both 
protocols, this distance increases significantly if Alice and Bob 
perform a few B-steps before switching to the one-way CSS-like EPP. 
More precisely, one application of a B-step already 
increases this maximum possible distance to approximately $30~{\rm km}$
in the four-state and to $34~{\rm km}$ in the six-state protocol.  
One may observe a sudden increase in the secret-key generation rate 
on applying a B-step. 
This is because a B-step decreases the bit-error rate significantly and thus 
the effect of dark counts becomes less significant. 
However, for larger distances, dark counts again become dominant, 
resulting in a new dip in the key generation rate unless a second 
B-step is applied. 
For increasing numbers of B-steps this effect becomes less dominating as the 
phase-error probability of the untagged pairs increases after each 
B-step \cite{LoGott,C02} and therefore dark counts become effective in 
the phase-error part. It can also be noticed that in the six-state protocol 
each B-step results in a larger increase of the maximal achievable 
distance with less reduction of the secret-key generation rate  
compared to the four-state protocol. Basically, this is due to the fact 
that the six-state protocol can sustain higher error rates.
Finally, as depicted in Fig. \ref{nvsdistbbsix}, multiple applications of 
B-steps quickly increase the maximum possible distances almost 
up to approximately $37~{\rm km}$ for the four-state and to $44~{\rm km}$ for 
the six-state protocol.

Let us now explore to which extent multiple applications of B-steps 
are capable of approaching the limiting distances resulting from
equation (\ref{MaxDistance}) for the two QKD protocols under consideration. 
These latter distances are indicated by full vertical lines in Figures 
\ref{BB84KTH1} and \ref{SixStateKTH1}.
Following the arguments of Ref. \cite{GLLP00} which form the basis for the 
secret-key 
generation rates of equations (\ref{GLLPRate}) and (\ref{BCSS}), for our 
purpose it is sufficient
to explore the possibility of purifying only the untagged qubit-pairs 
by B-steps. 
As demonstrated in Ref. \cite{Kedar}, the inequality
\begin{equation}
\left( q_{\rm I}-\frac{1}{4}\right) ^{2}+\left( q_{\rm z}-\frac{1}{4}\right) ^{2}>
\frac{1}{8},
\label{KedarLimit}
\end{equation}
is a necessary condition for the purification of a  
Bell-diagonal state of the form (\ref{tau}) by a sequence of B-steps 
followed by a CSS-like EPP. 
Therefore, using equations (\ref{inc1}-\ref{inc4})  
inequality (\ref{KedarLimit}) yields for the two protocols 
\begin{equation}
\Delta <\left\{
\begin{array}{l@{\quad \quad}l}
1-5\delta & \text{four-state protocol} \\
\frac{1}{2}\left( 2-5\delta -\sqrt{5}\delta \right) & \text{six-state protocol.}%
\end{array}%
\right.  \label{MaxDistEPP2}
\end{equation} 
The other solutions of the inequality (\ref{KedarLimit}) do not satisfy the 
necessary condition for secret-key generation given by (\ref{MaxDistance}).  

\begin{figure}
\vspace{0.7cm}\hspace{1.0cm}\resizebox{0.82\columnwidth}{!}{%
  \includegraphics{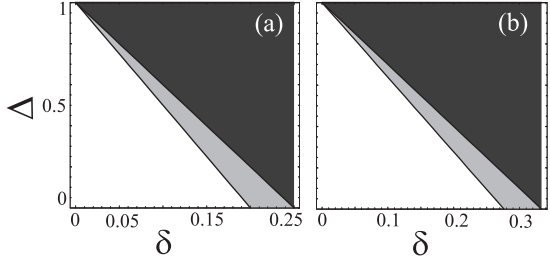}}
\caption{Regions bounded by equations (\ref{MaxDistance}) 
(white+grey) and (\ref{MaxDistEPP2}) (white) for the four-state (a) and
the six-state (b) QKD protocols. Secret key distillation is not possible 
by any means in the black region.
The grey  region is not accessible to B-steps and CSS-like EPP.}	
\label{regionsbb+6}
\end{figure}

In Figure \ref{regionsbb+6}, we plot the possible values of the 
effective tagging 
probability $\Delta$ and the estimated bit-error rate $\delta$ 
consistent with the necessary conditions (\ref{MaxDistance}) and 
(\ref{MaxDistEPP2}),  for the four- and the 
six-state protocols. These are basically parameters which can be 
estimated by Alice and Bob. 
According to the necessary condition (\ref{MaxDistance}), 
secret-key distillation is, in principle, possible everywhere 
except in the black regime. 
One may notice, however, the small grey region which (although 
allowed by inequality \ref{MaxDistance}) is not accessible to a 
post-processing involving B-steps and a CSS-like EPP. 
The upper bounds on distances resulting from 
the inequalities (\ref{MaxDistEPP2}) are indicated by the vertical 
dotted lines in Figures \ref{BB84KTH1} and \ref{SixStateKTH1}. 

From Figures \ref{BB84KTH1}, \ref{SixStateKTH1} and \ref{nvsdistbbsix}, 
it is apparent that these latter threshold values for the maximum possible 
distances are approached already after a few iterations of B-steps 
followed by a CSS-like EPP. 
Thus, such a combination yields a useful method for counteracting the 
influence of dark counts on the secret-key rate. 
However, from Figures \ref{BB84KTH1} and \ref{SixStateKTH1} it is 
also apparent that at the same time the secret-key rate decreases 
considerably with the application of more than two B-steps. 

As decoy-state protocols were developed in order to suppress imperfections 
arising from  multiphoton pulses
it is of interest to explore the influence of B-steps on the corresponding 
achievable secret-key generation rates. For this purpose let us consider
a decoy-state protocol involving two decoy WCPs with mean photon 
numbers $\kappa < \nu$ fulfilling the
additional requirement 
$\kappa~{\rm exp}(-\kappa) < \nu~{\rm exp}(-\nu)$, 
and a signal pulse with mean photon number $\mu > \kappa + \nu$. 
Therefore, the decoy pulses are detected with 
probabilities  $P^{(\kappa)}_{{\exp}}$ and $P^{(\nu)}_{{\exp}}$ 
obeying the relations \cite{Wang,MaQi05}
\begin{eqnarray}
P^{(\kappa)}_{{\exp}} &=& P_{{\exp}}^{\rm dark} e^{-\kappa} + 
s_1 \kappa e^{-\kappa} + s_{\rm m} (1 - e^{-\kappa} - 
\kappa e^{-\kappa}),\nonumber\\
P^{(\nu)}_{{\exp}} &\geq & P_{{\exp}}^{\rm dark} e^{-\nu} + s_1 \nu e^{-\nu} +
s_{\rm m} (1 - e^{-\kappa} - 
\kappa e^{-\kappa})\frac{\nu^2 e^{-\nu}}{\kappa^2 e^{-\kappa}}.
\nonumber\\
\label{decoy1}
\end{eqnarray}
Thereby, $s_{\rm m}$ is the conditional 
probability that the detector clicks provided a multiphoton 
pulse with mean photon number $\kappa$ hits the detector, whereas 
$s_1$ is the corresponding probability for single-photon pulses.
Using (\ref{decoy1}) we obtain 
\begin{eqnarray}
s_1 &\geq& \frac{\nu^2 e^{\kappa}P^{(\kappa)}_{\exp} - 
\kappa^2 e^{\nu} P^{(\nu)}_{\exp} - 
(\nu^2 - \kappa^2)P_{\exp}^{\rm dark}}{\kappa \nu (\nu - \kappa)}:=\overline{s}_1.
\end{eqnarray}
The inequality in the second line of (\ref{decoy1}) is valid 
provided the inequalities 
$\kappa < \nu$  and
$\kappa~{\rm exp}(-\kappa) < \nu~{\rm exp}(-\nu)$ are fulfilled. 
Correspondingly, the probability $\Delta_{\mu}$ of multiphoton signal pulses 
can be upper-bounded as follows 
\begin{eqnarray}
\Delta_{\mu} &\leq& 1 - \frac{\overline{s}_1 \mu e^{-\mu}
}{P_{\exp}^{(\mu)}}:=\tilde{\Delta}_{\mu}.
\end{eqnarray}

\begin{figure}
\vspace{0.7cm}\hspace{1.0cm}\resizebox{0.75\columnwidth}{!}{%
  \includegraphics{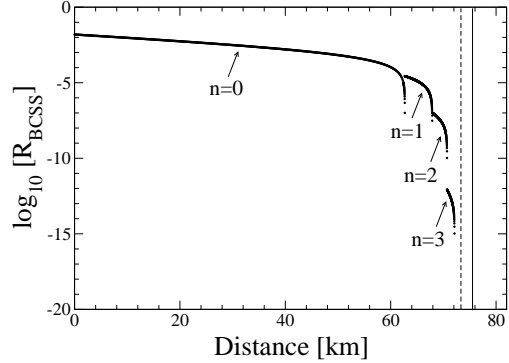}}
\caption{Four-state protocol with decoy pulses: The parameters are the 
same as in Figure \ref{BB84KTH1}, 
while $\mu=0.55$, $\kappa=0.10$, and $\nu=0.27$.}
\label{decoyBB84}
\end{figure}

\begin{figure}
\vspace{0.7cm}\hspace{1.0cm}\resizebox{0.75\columnwidth}{!}{%
  \includegraphics{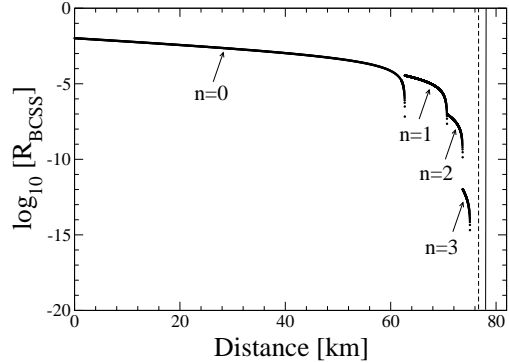}}
\caption{Six-state protocol with decoy pulses: The parameters are the 
same as in Figure \ref{decoyBB84}.}
\label{decoySixState}
\end{figure}

\begin{figure}
\vspace{0.7cm}\hspace{1.0cm}\resizebox{0.75\columnwidth}{!}{%
  \includegraphics{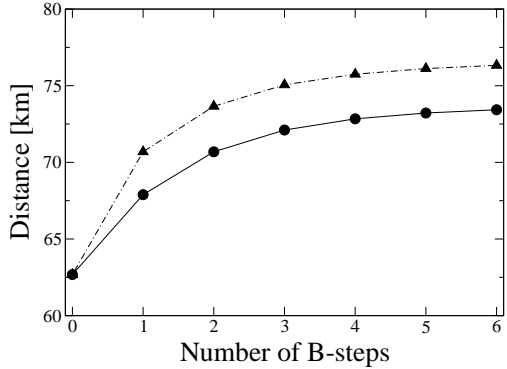}}
\caption{Maximum achievable distances for different numbers of B-steps for the
four-state (lower curve) and the six-state (upper curve) protocols with decoy pulses. }
\label{nvsdist1}
\end{figure}

To investigate the influence of B-steps on the achievable secret-key rates 
in the context of QKD  protocols with decoy pulses, we can adapt our previous 
arguments easily. 
In particular, a lower bound on the resulting secret-key generation rate is obtained from
equations (\ref{map}), (\ref{Pprime}), (\ref{Deltaprime}),  (\ref{Qprime}), and (\ref{BCSS}). 
Thereby, the recursive relations have to be solved by setting $\Delta=\tilde{\Delta}_\mu$ in the 
initial conditions (\ref{inc-six}) and (\ref{inc-four}) for the six- and the 
four-state protocol, 
respectively. These initial conditions take into account that the phase-error probability 
can be bounded from above by $\delta/(1-\tilde{\Delta}_\mu)$.  
The resulting lower bound on the secret-key generation rate and its dependence on the 
length of 
the optical fibre used for the transmission of photons
are depicted in Figures \ref{decoyBB84} and  \ref{decoySixState} for the four- and the six-state  
protocol, respectively. 
Following Ref. \cite{Wang}, we have chosen $\mu$, $\kappa$ and $\nu$ equal to 
0.55, 0.10 and 0.27, respectively.
Typically, multiple application of B-steps increase the distance over which a secret key can 
be exchanged significantly. 
The maximum distances and their dependence on the number of applied B-steps is shown in
Figure \ref{nvsdist1} for both protocols with decoy pulses.
The asymptotically achievable maximum distances 
of the order of $80~{\rm km}$ are reached already after
a few B-steps. Moreover, it is worth noting that the net increase in 
distance of about $15~{\rm km}$ (after 2 or 3 B-steps) is the same as that 
for the conventional four- and six-state protocols. 

In closing, we would like to stress once more that, although 
our results have been formulated in the framework of the 
two-way EPP-based versions of the four- and the six-state 
QKD protocols, they also apply to the corresponding 
pre\-pa\-re-and-measure schemes. 
According to Theorems 5 and 6 in Ref. \cite{LoGott}  
such a reduction is 
possible without compromising security, 
due to basic properties of the B-step and 
the one-way CSS-like EPP. Let us briefly highlight the 
main steps in this reduction. In the terminology of \cite{LoGott}, 
a B-step involves 
only measurements of the form ${\cal Z}_{\rm A}\otimes{\cal Z}_{\rm B}$ 
for the purpose of bit-flip error detection and subsequent rejection.  
Hence, no post-selection based on phase-error syndromes takes 
place during the two-way part of the post-processing. 
In the one-way  CSS-like part each operator being 
measured is either of ${\cal X}$- or ${\cal Z}$-type 
(definitions 1 and 4 of  Ref. \cite{LoGott}) while, at any rate,   
earlier measurements do not affect the sequence of subsequent 
measurements. Moreover, all the operators commute with each other and 
thus all the measurements of   
${\cal Z}$-type can be performed before all the 
measurements of ${\cal X}$-type. 
To complete the reduction, one has to note that ${\cal X}$-type 
measurements at the end of post-processing yield phase-error 
syndromes which do not affect the value of the final key and thus 
Alice and Bob do not have to perform them \cite{SP}. 
In the resulting prepare-and-measure schemes  
the classical post-processing involves 
a number of two-way error-rejection steps 
(parity checks) and additional one-way 
post-processing (error-correction and 
privacy amplification) based on asymmetric CSS codes. The phase-error syndrome 
measurements become effectively privacy amplification \cite{SP}. 

\section{Concluding remarks}
\label{Sec5} 
We have analyzed secret-key generation rates in the presence of 
imperfections arising from tagging of Alice's source and 
from dark counts at Bob's detectors. In particular, we considered a 
post-processing procedure (error correction and privacy amplification) 
based on a combination of B-steps and asymmetric CSS codes. 
As a main result, for the four-state, the six-state, and the decoy-state 
protocols, it was demonstrated that such a post-processing 
may considerably increase the maximum distances over which a 
secure key can be distributed in optical-fiber links. 

Hence, incorporation of B-steps in the post-processing stage of 
practical implementations of the protocols is proven to be 
particularly useful. The usefulness of B-steps is also 
one of the main results of a recent thorough investigation of 
decoy-state protocols presented in Ref. \cite{MDCTL06}. 
On the contrary, P-steps of the 
Gottesman-Lo type do not seem to be as useful as B-steps. 
Indeed, all our numerical simulations demonstrate 
that a few applications of B-steps are sufficient to bring 
the maximal secure distances very close to the upper bound. 
Recently, this inessentiality of P-steps has been pointed out 
by other authors as well \cite{Kedar,MDCTL06,ABBMM06}.

We would like to conclude this work with a discussion about 
certain assumptions underlying our approach and related possible 
open questions. Our model for sources and 
detectors is not as general as possible and it suffers from the same 
limitations as the model adopted in Refs. \cite{GLLP00,L00,BLMS00,FGSZ01}. 
For instance, we have focused on imperfect sources which tag 
a fraction of the signals in an uncorrelated and independent manner.  
In other words, we have not considered the case of coherent  
or other highly correlated basis-dependent tagging \cite{GLLP00}. 
In this context, 
our analysis has been based on a particular class 
of eavesdropping attacks where the sets of tagged and 
untagged qubits are attacked separately. Each tagged qubit is treated 
individually (by means of a PNS attack) whereas untagged 
qubits undergo a coherent (joint) attack. In this way, on the one hand we 
essentially give Eve a perfect copy of the part of the key originating 
from tagged signals, while on the other hand 
we give her all the power to retrieve as much information as 
possible about the remaining bits of the key. It is plausible that, 
for a fixed bit-error rate, this is the most powerful attack 
one may consider in the framework of the particular model for sources and 
detectors. However, we would like to emphasize that this work shows 
how incorporation 
of B-steps prior to one-way CSS-based post-processing can postpone certain 
dark-count effects  thus  increasing the distances over which a 
secret-key can be distributed.  This result is 
quite general and is not expected to change in the case of 
other, perhaps more efficient and more powerful, eavesdropping 
strategies. Indeed, as pointed out in Refs. \cite{L00,FGSZ01}, the maximum 
secure distances for WCP-based QKD are not limited by the eavesdropping 
strategy under consideration but rather by the actual 
detector performance and especially by the dark-count rate.
Finally, throughout this work we have also not addressed the case of 
imperfect sources which emit weak coherent pulses 
with nonrandom phases or highly dimensional signals \cite{GLLP00}. 
At any rate, all of these issues depend on how well 
the two legitimate users know their devices (e.g., source and detectors)  
and how reliably they can characterize them   
by means of other, perhaps untrusted, apparatus \cite{MayYao}.  

\section{Acknowledgments}
This work is supported by the EU within the IP SECOQC. 
Informative discussions with N. L\"utkenhaus, K. S. Ranade, and J. M. 
Renes are acknowledged. 
GMN also acknowledges support by ``Pythagoras II'' of the EPEAEK 
research programme.


\begin{thebibliography}{99}
\bibitem{note2}
Classical random sampling theory can be applied safely for the
estimation of error rates and the establishment of related
confidence levels during the verification test (see, for instance,
Ref. \cite{LCA} for details). For the sake of simplicity, throughout
this work we assume that the bit-error rate estimated during the
tests is the actual bit-error rate in the pairs shared between Alice
and Bob
\end{thebibliography}
\end{document}